\documentclass[aps,prd,showpacs,showkeys,preprintnumbers]{revtex4}
\usepackage{graphicx}                                 
\usepackage{amsmath,amsfonts}
\graphicspath{{./Figures/}}                           

\usepackage{epsfig}
\usepackage{amssymb}
\usepackage{amsfonts}
\usepackage{float}
\newcommand{\bc}{\texttt{$\;\mathrm{bc}\;$}}                         
\newcommand{\fc}{\texttt{$\;\mathrm{fc}\;$}}                         

\newcommand{\beq}{\begin{equation}}
\newcommand{\eeq}{\end{equation}}
\newcommand{\bea}{\begin{eqnarray}}
\newcommand{\eea}{\end{eqnarray}}
\newcommand{\beas}{\begin{eqnarray*}}
\newcommand{\eeas}{\end{eqnarray*}}
\newcommand{\eq}{\begin{equation}}
\newcommand{\en}{\end{equation}}
\newcommand{\eqa}{\begin{eqnarray}}
\newcommand{\ena}{\end{eqnarray}}

\begin{document}

\preprint{HU-EP-07/22, ITEP-LAT/2007-13}

\title{Improved Landau gauge fixing and the suppression of 
finite-volume effects of the lattice gluon propagator}
\author{I.L.~Bogolubsky}
\affiliation{Joint Institute for Nuclear Research, 141980 Dubna,
Russia}
\author{V.G.~Bornyakov}
\affiliation{Institute for High Energy Physics, 142281 Protvino, Russia}
\affiliation{Institute of Theoretical and Experimental Physics,
Moscow, Russia}
\author{G.~Burgio}
\affiliation{Universit\"at T\"ubingen, Institut f\"ur Theoretische Physik,
72076 T\"ubingen, Germany}
\author{E.-M.~Ilgenfritz}
\affiliation{Humboldt-Universit\"at zu Berlin, Institut f\"ur Physik,
12489 Berlin, Germany}
\author{V.K.~Mitrjushkin}
\affiliation{Joint Institute for Nuclear Research,
141980 Dubna, Russia}
\affiliation{Institute of Theoretical and Experimental Physics,
Moscow, Russia}
\author{M.~M\"uller--Preussker}
\affiliation{Humboldt-Universit\"at zu Berlin, Institut f\"ur Physik,
12489 Berlin, Germany}

\date{January 9, 2008}

\begin{abstract}
For the gluon propagator of pure $SU(2)$ lattice gauge theory in the
Landau gauge we investigate the effect of Gribov copies and finite-volume
effects. Concerning gauge fixing, we enlarge the accessible gauge orbits by adding
nonperiodic $\mathbb{Z}(2)$ gauge transformations and systematically employ
the simulated annealing algorithm. Strategies to keep all $\mathbb{Z}(2)$ sectors
under control within reasonable CPU time are discussed. We demonstrate that the
finite-volume effects in the infrared regime become ameliorated.
Reaching a physical volume of about $(6.5 {\rm~fm})^4$, we find that the
propagator, calculated with the indicated improvements, becomes flat in the
region of smallest momenta. There are first signs in four dimensions of a 
decrease towards vanishing momentum.

\end{abstract}

\keywords{Landau gauge, Gribov problem, simulated annealing, gluon propagator}
\pacs{11.15.Ha, 12.38.Gc, 12.38.Aw}
\maketitle

\section{Introduction}
In recent years the infrared behavior of gauge-variant Green's functions
of Yang-Mills theories has increasingly attracted interest. This fact is mainly
related to the existence of the Landau (or Coulomb) gauge confinement scenarios
proposed by Gribov~\cite{Gribov:1977wm} and Zwanziger~\cite{Zwanziger:1993dh}
on one hand and by Kugo and Ojima~\cite{Kugo:1979gm} on the other. The
interest was stimulated by the practical progress achieved over the years within
the Dyson-Schwinger equation (DSE) approach as pursued by Alkofer, von Smekal and
others (for an intermediate review see~\cite{Alkofer:2000wg}). Lattice gauge
theory is able to check these scenarios from first principles. For example, one
can compare lattice results with analytic and numerical solutions of the
(truncated) hierarchy of DSE, however within the limitations of finite lattice
disretisation and  - even more important in this respect - of finite-volume
effects. One crucial test concerns the proposed infrared vanishing (diverging)
of the Landau gauge gluon (ghost) propagator. The closely related behavior of
the two propagators is intimately connected with an infrared fixed 
point~\cite{von_Smekal:1997is,von_Smekal:1997vx} of the momentum subtraction (MOM) 
scheme~\cite{Chetyrkin:2000fd} running QCD coupling 
(see also e.g. \cite{Shirkov:2002gw}).  So far,
only in two and three dimensions it was possible to reach the expected asymptotics
in an unambigious manner~\cite{Maas:2006qw,Cucchieri:2007uj,Maas:2007uv}.
In four dimensions, for $SU(2)$ as well as $SU(3)$ lattice gauge theory, the
ultimate decrease of the gluon propagator towards vanishing momentum has not
yet been established. This paper is devoted to this question but restricted to
the $SU(2)$ case.

A possible pattern of finite-volume deviations from the far-infrared behavior
of the gluon and ghost propagators has been pointed out thanks to the 
formulation and solution of the DSE in a compact space-time~\cite{Fischer:2007pf}. 
The sobering message is that really infrared results can be expected only on 
lattices of linear sizes $L=O(10 {\rm~fm})$. 
However, in the DSE approach the Gribov ambiguity is assumed not to play 
a relevant role, such that something comparable about the gauge-fixing
vulnerability of the propagators cannot be learned from DSE solutions.
Nevertheless, the restriction to the {\it fundamental modular region} 
might also considerably change the structure of the DSE at finite 
volume~\cite{Zwanziger:1993dh}.

In the present paper we study the question to what extent the finite-volume
effects observed in lattice calculations can be related to the
existence of Gribov copies and can be cured (for presently accessible volumes)
by a better treatment of the Gribov ambiguity,
i.e. systematically pursuing a restriction to the fundamental modular region.
The common hope
is that in the limit of infinitely large volume Gribov copy effects become
negligible. If this is true, then
the random choice of an arbitrary gauge copy in the Gribov region (which is
statistically equivalent to an average over all of them) should be the
physically adequate solution~\cite{Zwanziger:2003cf}.

In paper~\cite{Bogolubsky:2005wf} it has been noted that
enlarging the gauge orbits by nonperiodic $\mathbb{Z}(2)$ gauge
transformations (called  ``$\mathbb{Z}(2)$ flips'') generically leads to larger 
values of the gauge functional $F$. In this paper we continue to explore this 
approach. Furthermore, within the traditional, continuous part of the 
gauge-fixing problem, we systematically employ the simulated annealing algorithm.
Testing these two modifications, we find that in the range of linear lattice sizes
between $L \simeq 2 {\rm~fm}$ and $6.5 {\rm~fm}$ the choice among Gribov copies, 
and therefore the optimization of the gauge-fixing method, is still important.
Our paper represents a systematic extension of the previous work, 
where the $\mathbb{Z}(2)$ flips have been
studied for the first time~\cite{Bogolubsky:2005wf}. Besides being much less 
volume dependent, the gluon propagator in the extended Landau gauge is found
flattened for momenta $p < 0.5 {\rm~GeV}$, 
and there are first indications for a decrease towards the infrared limit.

Section II will give an introduction to the necessary technical details.
In Sec. III we discuss steps towards an optimal gauge-fixing strategy.
The Gribov copy effects at finite volumes are pointed out in Sec. IV.
In Sec. V all our results, obtained on various lattices with the respective
optimal strategy, are put together and we summarize our findings.

\section{General setup: extension of the Landau gauge}
Like many other investigators of the SU(2) gluon propagator we compute it
with Monte Carlo (MC) techniques on a lattice with periodic boundary conditions.
The standard Wilson single-plaquette action and the lattice definition for
the gauge potentials
\beq
A_{\mu}(x+\hat{\mu}/2) = A^b_{\mu}(x+\hat{\mu}/2)~\frac{\sigma^b}{2} = \frac{1}{2iag_0} \left(U_{x\mu} - U^{\dagger}_{x\mu}\right)
\label{gauge_potential}
\eeq
are adopted. In order to fix the Landau gauge for each lattice gauge field
$\{U\}$ generated by means of a MC procedure, the gauge functional
\beq
F[g]= \frac{1}{2} \sum_{x,\mu}
      \mathrm{tr} \left( g(x) U_{x\mu} g^{\dagger}(x+\hat{\mu}) \right)
\label{gauge_functional}
\eeq
is iteratively maximized with respect to a gauge transformation $~g(x)~$
which is usually taken as a periodic field, too.

In order to approach the global maximum (related to the fundamental modular
region) as close as possible, we are using the simulated annealing (SA)
algorithm~\cite{Kirkpatrick:1983aa},
in combination with subsequent standard overrelaxation (OR). The latter
is applied in the final stage of the gauge-fixing procedure in order to finalize
the transformation to any required precision of the transversality
condition $~\partial_{\mu} A_{\mu} = 0$.
A decade ago, SA has been shown to be very efficient, when dealing with
the maximally Abelian gauge (MAG)~\cite{Bali:1994jg,Bali:1996dm}.
In the latter case typically a huge number of local extrema of the gauge functional
is observed. The effectiveness of the SA algorithm in the case of the Landau gauge
remained quite unclear for a long time. It was practically used for this
gauge in the first study of the ghost propagator~\cite{Suman:1995zg}.
In Ref.~\cite{Gutbrod:1996sq}, also for the Landau gauge, a comparison with other
algorithms was carried out. This comparative study came to the conclusion that
SA might not provide a real advantage. Today, the state of the art is that SA
is practiced in a hybrid form, mixed with microcanonical update steps. It is
repeatedly started from random gauge transformations $~g(x)~$ and ends with
OR, producing each time one gauge copy in the Gribov region. In a recent,
more thorough investigation~\cite{Schemel:2006xx} this version of the
SA algorithm was seen to become superior,
with growing lattice size, to the repeated application of
the pure OR algorithm. The efficiency was quantified by the ability to
produce a better (narrower) distribution of copies (local extrema) within less
or equal CPU time.  The results of this study will be published
elsewhere~\cite{Schemel:2007xx}.

The SA algorithm, in the present context, generates a field of gauge transformations
$~g(x)~$ by MC iterations with a statistical weight proportional to 
$~\exp{(F[g]/T)}~$. The ``temperature'' $~T~$ is a technical parameter which is 
gradually decreased in order to maximize the gauge functional $F[g]$. In the 
beginning, $~T~$ has to be chosen sufficiently large in order to allow traversing 
the configuration space of $~g(x)~$ fields in large steps. It has been checked 
that an initial value $~T_{\rm init}=1.5~$ is high enough. After each
quasiequilibrium sweep, including both heatbath and microcanonical updates,
$~T~$ has been decreased with equal step size until $~g(x)~$ is uniquely
captured in one basin of attraction. The criterion of success is that
during the following OR the violation of transversality decreases in a
monotonous manner for almost all applications of the compound algorithm.
This condition is reasonably satisfied for a final lower temperature value
$~T_{\rm final}=0.01~$~\cite{Schemel:2006xx}. The number of temperature steps
was chosen of the order $O(10^3)$.

The second novel feature of our gauge-fixing procedure compared to the standard
ones is the application of $\mathbb{Z}(2)$ flip transformations, the essence of
which is an extension of the gauge orbits for any MC generated lattice
configuration. We will abbreviate the extended gauge-fixing method as the FSA
(flip-SA) algorithm. There is room for its realization under various strategies
(see below) that can be chosen in order to save computing time. The flip
transformation was first considered in the context of Landau gauge fixing in
Ref.~\cite{Bogolubsky:2005wf}. For $SU(2)$ gauge theory, each flip transformation
consists of a simultaneous $\mathbb{Z}(2)$ flip of all links
$~U_{\nu}(x) \to - ~U_{\nu}(x)~$ throughout a 3D hyperplane at a given value of
the coordinate $~x_{\nu}$. This is just a particular case of a gauge transformation
which is not periodic but periodic modulo $\mathbb{Z}(2)$,
\beq
g(x+L\hat{\nu}) = z_{\nu} g(x)\,, \qquad z_{\nu}=\pm 1 \in \mathbb{Z}(2) \, .
\eeq

It is obvious that the above transformation of the gauge field leaves the gauge
field action as well as the path integral measure invariant
(note that this symmetry is unbroken in the confinement phase).
This would not be true anymore in a gauge theory with a
fundamental matter field. Therefore, the $\mathbb{Z}(2)$ flip transformation
cannot be applied to such models.

With respect to the flip transformation all gauge copies of one given
field configuration relative to the initial gauge can be split into
$~2^4=16$ sectors for $SU(2)$ gauge fields ($3^4=81$ sectors for $SU(3)$).
Within each of these sectors - all being present in the path integral measure -
different gauge copies are connected by continuous, strictly periodic gauge
transformations. With this new element, our gauge-fixing  procedure consists
of two steps: the first one is to choose the best out of the $16$ flip sectors
and the second one with the help of SA is to find the gauge copy with the highest
value of the gauge functional while staying within the given sector. In practice,
both steps are performed in an intertwined manner, because the decision which is
the ``best'' sector in principle requires knowing the best copy of each sector.
It is immediately clear that this procedure allows to find higher local maxima
of the gauge functional (\ref{gauge_functional}) than the traditional
gauge-fixing procedures. The latter by default choose for a given configuration 
only one flip sector, and in most of the cases only one copy in this sector.
The sector taken is usually the one
randomly selected by the MC update algorithm. It is equivalent to averaging
over all flip sectors and therein over copies within the so-called Gribov region.

Obviously the two prescriptions to fix the Landau gauge, the traditional one
and the new one, are not equivalent. Indeed, for some modest lattice
volumes it has been shown in Ref.~\cite{Bogolubsky:2005wf}
that they give rise to different results for the gluon as well as the ghost
propagators. In the present paper for the gluon propagator we want to present
some numerical evidence that the results converge to each other in the large
volume limit. The ghost propagator under this extended Landau gauge fixing will
be addressed in a future publication.

The computations presented in this work have been done at rather strong coupling,
at $\beta\equiv 4/g_0^2 = 2.20$ . The reason for this choice was to get access
to a comparatively large physical volume. We fix the scale taking the string 
tension as $\sigma$ = (440 MeV)$^2$ and adopting the lattice value
$\sqrt{\sigma} a = .469$ found in Ref.~\cite{Fingberg:1992ju}.
Thus, our largest lattice size $32^4$ has a physical size of about
$(6.5 {\rm~fm})^4$. In order to study the volume dependence we have calculated
the gluon propagator also for smaller lattices, such that we have sizes
ranging from $L^4=8^4$ to $32^4$.

\section{The quest for an optimal gauge-fixing strategy}
As a first step we have searched for an optimal strategy to find the best
gauge copy for each lattice size. On $16^4$ (and $24^4$) lattices we have
produced ensembles of $60$ ($46$) MC configurations. For each configuration
we created with the help of SA $5$ gauge copies as local maxima of the
gauge functional $~F~$ within each of the $16$ flip sectors, i.e. in total $80$
gauge copies per MC field configuration. In a production run we would like
to get along with considerably less copies per MC configuration.
This will become particularly important for $SU(3)$, where one has
to deal with $3^4=81$ different $\mathbb{Z}(3)$ sectors.

By $~\langle F_{ns}(nc) \rangle~$ let us denote the MC ensemble average
over the maximized functional values $~F~$ taken from
all $16$ sectors ($~ns = 16~$) or a random subset of $~ns < 16~$ flip sectors
and from the best of $~nc \le 5~$ gauge-fixed copies.
These copies are created sequentially, starting from new random periodic copies,
in each of the $ns$ chosen sectors and the best one is stored.
The average $\langle F_{16}(5)\rangle~$ corresponds to the
largest accessible (best) functional values. Representing the largest
affordable computing effort it will serve as a reference value.
Table \ref{tab:gaugefunctional} shows the values for the different cases.
One sees that the functional values become larger, when all 16 flip
sectors are taken into account. The data clearly indicate that (for the given 
volume) it is more important to scan all 16 sectors than to search for the 
best copy in one (randomly chosen) sector.
But the improvement is much less dramatic
for the larger lattice size $24^4$ than for the $16^4$ lattice. The
reference values for the functional are very close for the two lattice
sizes in contrast to the cases $ns=1$ of one randomly chosen flip sector.
Moreover, we see that 5 random copies already seem to be optimal for both the
lattice sizes.
\begin{table*}
\begin{center}
\mbox{
\begin{tabular}{|c|c|c|c|}\hline
       &      & $\langle F_{ns}(nc) - F_0  \rangle$ & 
       $\langle F_{ns}(nc) - F_0  \rangle$ \\ 
       $ns$ & $nc$ & for $16^4$ & for $24^4$ \\
\hline\hline
 1 & 1 & $  1(8) \cdot 10^{-5}$ & $ 25(4) \cdot 10^{-5}$  \\ \hline
 1 & 5 & $  6(8) \cdot 10^{-5}$ & $ 31(4) \cdot 10^{-5}$  \\ \hline\hline
16 & 1 & $ 32(9) \cdot 10^{-5}$ & $ 36(4) \cdot 10^{-5}$  \\ \hline
16 & 2 & $ 33(9) \cdot 10^{-5}$ & $ 38(4) \cdot 10^{-5}$  \\ \hline
16 & 3 & $ 34(9) \cdot 10^{-5}$ & $ 38(4) \cdot 10^{-5}$  \\ \hline
16 & 4 & $ 34(9) \cdot 10^{-5}$ & $ 39(4) \cdot 10^{-5}$  \\ \hline
16 & 5 & $ 34(9) \cdot 10^{-5}$ & $ 39(4) \cdot 10^{-5}$  \\ \hline
\end{tabular}
}
\end{center}
\caption{The average gauge functionals $\langle F_{ns}(nc)\rangle$ as explained in
the text and subtracted with $F_0=0.82800$. For the lattice sizes $16^4$ and $24^4$
the numbers of investigated MC configurations are $60$ and $46$, respectively.
The inverse coupling is $\beta=4/g_0^2=2.20$.
}
\label{tab:gaugefunctional}
\end{table*}
In Table \ref{tab:diff_gaugefunctional} we show additionally the deviations or
distances
$\Delta_{ns,ns^{\prime}}(nc,nc^{\prime})=
                    \langle F_{ns}(nc)-F_{ns^{\prime}}(nc^{\prime}) \rangle$
between would-be runs with different numbers $ns$ and $nc$. The $\Delta$-values
have quite small statistical errors since the differences are always computed
configuration by configuration.
\begin{table*}
\begin{center}
\mbox{
\begin{tabular}{|c|c|c|c|c|c|c|}   \hline
 & & & & &
$\Delta_{ns,ns^{\prime}}(nc,nc^{\prime})$ & $\Delta_{ns,ns^{\prime}}(nc,nc^{\prime})$ \\
    & $ns$ & $nc$ & $ns^{\prime}$ & $nc^{\prime}$ & for $16^4$ & for $24^4$ \\
\hline\hline
A & 16 & 5 &  1 & 5 & $2.8(1) \cdot 10^{-4}$ & $8.5(4) \cdot 10^{-5}$  \\ \hline\hline
B &  1 & 5 &  1 & 1 & $4.6(4) \cdot 10^{-5}$ & $5.3(2) \cdot 10^{-5}$  \\ \hline\hline
C & 16 & 5 & 16 & 1 & $1.3(1) \cdot 10^{-5}$ & $2.9(2) \cdot 10^{-5}$  \\ \hline
D & 16 & 5 & 16 & 2 & $4.9(8) \cdot 10^{-6}$ & $1.5(1) \cdot 10^{-5}$  \\ \hline
E & 16 & 5 & 16 & 3 & $2.5(5) \cdot 10^{-6}$ & $7.6(7) \cdot 10^{-6}$  \\ \hline
\end{tabular}
}
\end{center}
\caption{Distances
$\Delta_{ns,ns^{\prime}}(nc,nc^{\prime})=
                    \langle F_{ns}(nc)-F_{ns^{\prime}}(nc^{\prime}) \rangle$
as defined in the text.
The statistics and the inverse coupling are the same as quoted in
Table \ref{tab:gaugefunctional}.
}
\label{tab:diff_gaugefunctional}
\end{table*}

From this work as well as from our earlier experience we know that the
functional $F$ and the gluon propagator at small momenta  are
anticorrelated (more detailed description of this anticorrelation
will be published elsewhere). We wish to emphasize that substantial
decrease of $\Delta$ with increasing volume shown in Table II does {\it not}
imply that the effect of improved gauge fixing on the propagator decreases
also that much.

Notice that the values in Table \ref{tab:diff_gaugefunctional} fall monotonously
from comparison A to comparison E for both the lattice sizes.
For $24^4$ the variation covers only one order of
magnitude compared with two orders for $16^4$. The variation of the best copy
results ($nc,nc'=5$) comparing the best sector ($ns=16$) with the first random sector
($ns^{\prime}=1$) (comparison A) shows the sectors to differ much 
more strongly from each
other on the smaller lattice than on the larger one. This indicates that,
concerning the gauge functional, the r\^ole of the flip sectors is
weakening with increasing volume. On the other hand the variation
between different copies within the same random flip sectors (case B) or within
the best sectors (C, D, E) becomes stronger the larger the lattice is. Therefore,
in order to distinguish the best sector we certainly need to generate more
gauge copies per sector the larger the lattice volume is. How many copies are
required within a given sector depends on the deviation from the reference value
one considers to be tolerable (compare with cases C, D, E).

These observations suggest a strategy to keep the total number of gauge copies
as low as possible, that have to be generated in order to guarantee a
certain prescribed closeness of the {\it average best gauge functional} to the
reference case. Since for the smaller lattice sizes
the functional values of $~F~$ for different gauge copies generated within
the best sector are scattered very closely to the maximal value in that sector,
we try to identify the best sector by gauge-fixing not more than one gauge copy
per sector. Actually this becomes difficult or even impossible for a larger volume.
After the best sector has been figured out, we could generate a few more gauge
copies for this particular sector only. In order to increase the probability not
to misidentify the best sector, compared to making only one gauge-fixing attempt
in all sectors, it is reasonable to perform a few more gauge fixings in a few
sectors that have already been recognized as good pretenders of being the best
sector.

In shorthand, we denote as ``$16 + 4$'' a strategy, where we first fix
one gauge copy in all $16$ sectors, and then fix a second, independent copy in
the $4$ best-candidate sectors, those with the highest ranking gauge functional
values of the gauge copy found first.
Taking again our data for the $24^4$ lattice with $80$ gauge copies per
configuration as the reference case to compare with, we checked the reliability
of such an improved strategy. We get a difference
$\langle F_{16}(5)-F_{16+4} \rangle = 1.9(2) \cdot 10^{-5}$, i.e.
almost the closeness to the reference case that was obtained with two gauge
copies in all sectors, although now, in the ``$16 + 4$'' strategy, a second copy
has been fixed in only $4$ out of $16$ sectors. As a compromise between the quality
and the need to limit the CPU time we have in practice chosen a strategy with
``$16 + 4 * 2$'' copies, i.e. in four selected sectors not one but two more
gauge copies are created. On our test ensemble of $46$ primary Monte Carlo
configurations we get a difference from the reference value
$\langle F_{16}(5)-F_{16+4*2} \rangle = 1.4(2) \cdot 10^{-5}$.

We have attempted to apply the same ``$16 + 4 * 2$'' strategy to $32^4$ lattices
as well. We have observed that for this lattice size the best sectors are not
so clearly distinguishable from the other sectors with generically lower values of
the gauge functional. For this reason we decided to produce additionally 16 copies,
one per sector. Thus we generated in total $40$ copies per MC configurations on
this lattice (``$16 * 2 + 4*2$''), instead of $80$.

For $12^4$ lattices we have blindly generated $5$ copies in each sector
(``$16 * 5$''), and for $8^4$ lattices just $3$ copies in each sector
(``$16 * 3$''). We found confirmation of the features observed for $16^4$ and
$24^4$ lattices as discussed in the beginning of this section.

Our produced ensembles of gauge-fixed field configurations are quoted in
Table~\ref{tab:statistics} together with the strategy used in each case.
$\langle F^{bc} \rangle$ is the average gauge functional for the best copy
(\bc) found by means of the preferential strategy at the given lattice size.
The difference $\langle F^{bc} - F^{fc} \rangle$
means the difference between the values achieved with the preferential
strategy (based always on access to all $16$ sectors) and the value found
for the first copy (\fc), i.e. for just one randomly chosen flip sector
and one copy. For comparison also
some values obtained with the standard OR method with $ns=1$ (i.e. no flips)
and $nc=1$ (one copy) are shown. The statistics for the OR procedure was
generally smaller but of the same order of magnitude as shown in the
second column of Table \ref{tab:statistics}.

\begin{table*}
\begin{center}
\mbox{
\begin{tabular}{|c|c|c|c|c|c|}   \hline
$L$ & $\#$ & strategy & $\langle F^{bc} \rangle $ & $ \langle F^{bc} - F^{fc} \rangle $
  & $\langle F^{fc}_{OR} \rangle$\\ \hline
 8  & 200 & ``$16 * 3$''          & 0.82721(23)  & 0.00298(7)  & 0.82365(25) \\ \hline
 12 & 200 & ``$16 * 5$''          & 0.82817(10)  & 0.00077(2)  & 0.82715(11) \\ \hline
 16 &  60 & ``$16 * 5$''          & 0.82834(9)   & 0.00028(1)  &             \\ \hline
 16 & 180 & ``$16 + 4 * 2$''      & 0.82834(8)   & 0.000244(6) & 0.82779(5)  \\ \hline
 24 &  46 & ``$16 * 5$''          & 0.82839(4)   & 0.000085(4) &             \\ \hline
 24 & 300 & ``$16 + 4 * 2$''      & 0.82843(2)   & 0.000132(2) & 0.82805(3)  \\ \hline
 32 & 247 & ``$16 * 2 + 4 * 2$''  & 0.82843(1)   & 0.000075(1) & 0.82815(1)  \\ \hline
\end{tabular}
}
\end{center}
\caption{Lattice sizes, statistics, gauge-fixing strategy employed and
the data on  average values of the gauge functional $ F $.
The meaning of $F^{bc}$, $F^{fc}$ and of $F^{fc}_{OR}$ is explained in the text.}
\label{tab:statistics}
\end{table*}

\section{The gluon propagator: Gribov copy and finite-volume effects}
The gluon propagator is defined by
\beq
D_{\mu\nu}^{ab}(p)=\langle \widetilde{A}_{\mu}^a(k) \widetilde{A}_{\nu}^b(-k) \rangle
                  =\left( \delta_{\mu\nu} - \frac{p_{\mu}~p_{\nu}}{p^2} \right)
            \delta^{ab} D(p)\,,
\label{gluonpropagator}
\eeq
where $\widetilde{A}(k)$ represents the Fourier transform of the gauge potentials
according to Eq. (\ref{gauge_potential}) after having fixed the gauge. The momentum
$p$ is given by $p_{\mu}=(2/a) \sin{(\pi k_{\mu}/L)}, ~~k_{\mu} \in (-L/2,L/2]$.
For $p \ne 0$, one gets
\beq
D(p) = \frac{1}{9} \sum_{a=1}^3 \sum_{\mu=1}^4 D^{aa}_{\mu\mu}(p) \; ,
\eeq
whereas at $p = 0$ the ``zero momentum propagator'' $D(0)$ is defined as
\beq
D(0) = \frac{1}{12} \sum_{a=1}^3 \sum_{\mu=1}^4 D^{aa}_{\mu\mu}(p=0) \; .
\eeq

\begin{figure*}
\vspace*{0.8cm}
\includegraphics[width=0.6\textwidth]{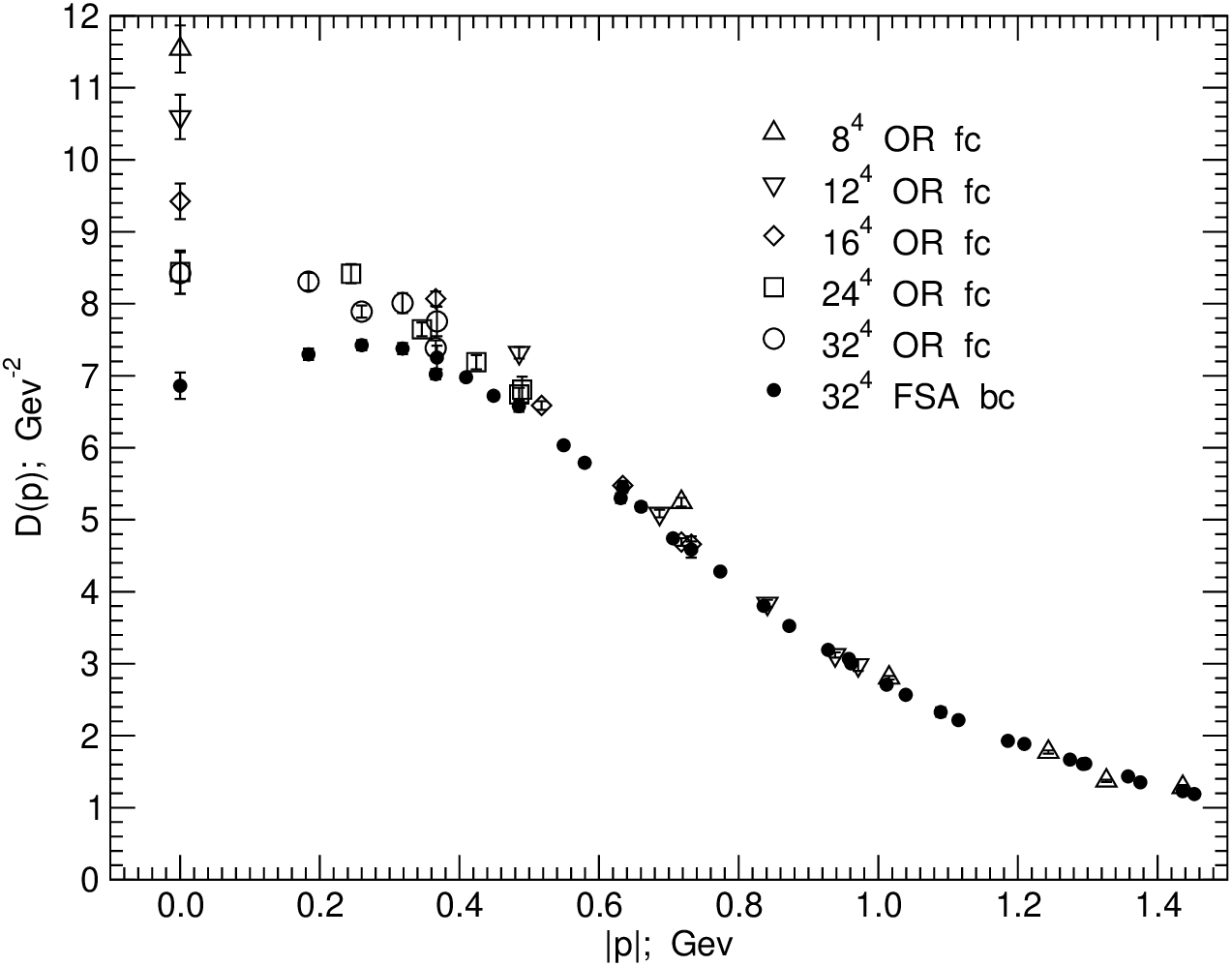}
\caption{The lattice gluon propagator versus momentum for $\beta=2.20$ and
various lattice sizes obtained by OR in comparison with FSA results for $32^4$.}
\label{fig:SA_vs_OR}
\end{figure*}

In order to compare with standard methods employed by other authors we have carried
out our own analysis with standard overrelaxation (OR) without $\mathbb{Z}(2)$ flips 
and restricting always to the first gauge copy. The corresponding findings together 
with our  \bc--FSA results obtained with the ``$16*2+4*2$'' strategy on the largest 
lattice $32^4$ are plotted in Fig.~\ref{fig:SA_vs_OR}.
We have convinced ourselves that the OR results for the $24^4$ lattice are in perfect
agreement with those recently obtained for a $22^4$ lattice and the same $\beta=2.20$
in Ref.~\cite{Cucchieri:2007uj}. A deviation of our FSA results in the infrared
($p < 0.4 {\rm~GeV}$) towards lower values of $D(p)$ becomes clearly visible.

As one might expect, due to the bias towards a larger gauge functional
in the case of the FSA algorithm (compared with the OR algorithm)
not only the expectation value of the gluon propagator becomes suppressed
at low momenta, but also the statistical fluctuations of the gluon propagator
become reduced. The effect is most clearly seen for the zero momentum
propagator $D(0)$. 

\begin{figure*}
\vspace*{1cm}
\includegraphics[width=0.6\textwidth]{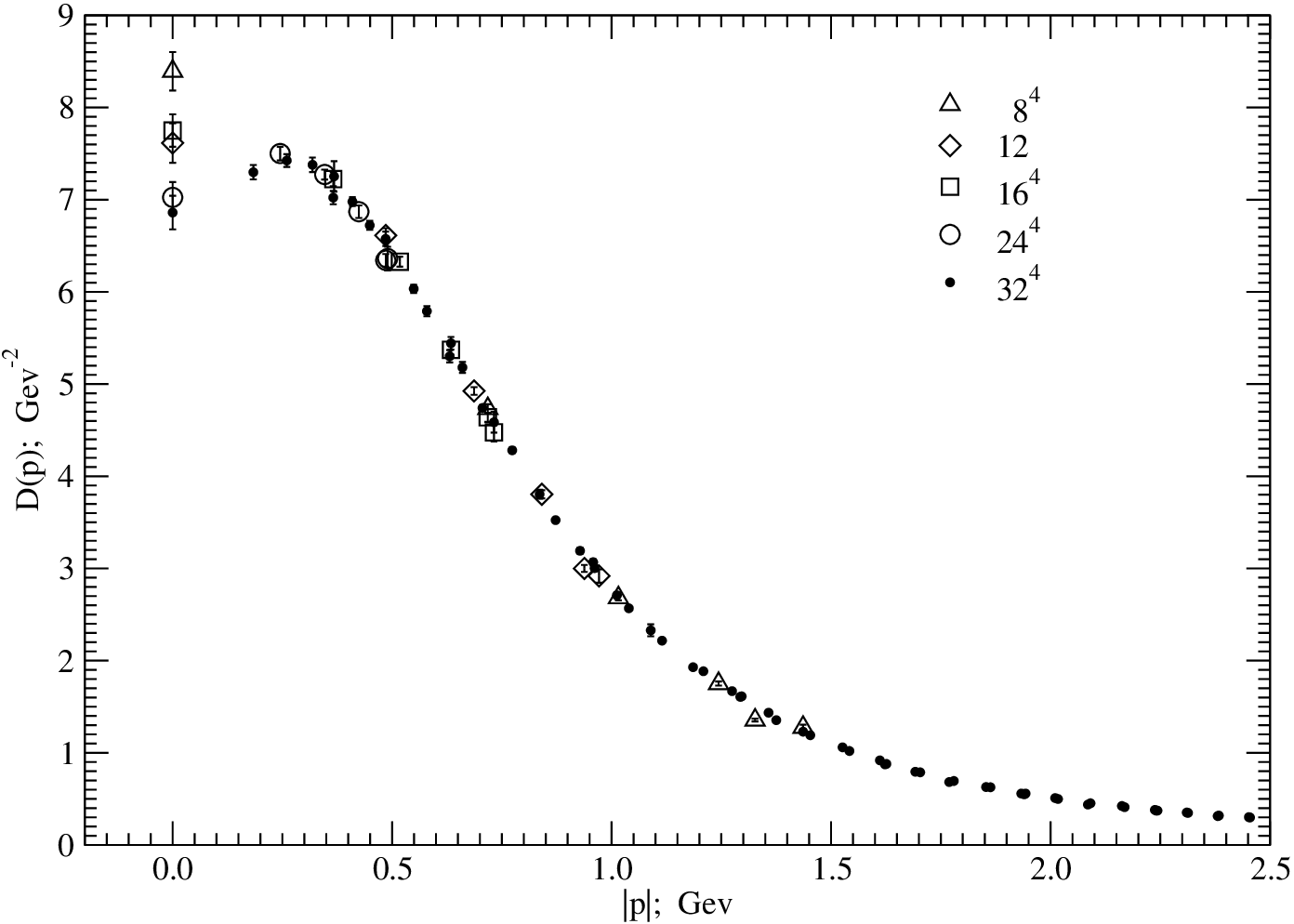}
\caption{The gluon propagator obtained with FSA gauge fixing in the infrared region
for various lattice sizes, all simulated at $\beta=2.20$.}
\label{fig:Gl_main}
\end{figure*}

In comparison to the finite-size dependence showing up after gauge fixing
with standard OR our new FSA method provides results very stable against varying
lattice size. This is demonstrated in Fig.~\ref{fig:Gl_main} collecting our
main results. All data points nicely fall onto a universal curve.
Indeed, comparing the data for different lattice sizes entering
Fig.~\ref{fig:Gl_main}  one can see that the finite-volume effects for the
momenta shown in the figure are indeed small. This is particularly important
for the minimal nonzero (on-axis) momenta for each given lattice size which
are {\it not excluded} from the plot.
Notice that for lattices $24^4$ and $32^4$ all momenta with components
$k_{\mu}$ satisfying the condition~\cite{Leinweber:1998uu}
\beq
\sum_{\mu} k_{\mu}^2 - \left(\sum_{\mu} \frac{1}{2} k_{\mu} \right)^2 < 3
\eeq
are shown. 
There is no significant breaking of rotational invariance for the momenta
included in the figure. Only in the case of OR there is one bigger deviation  
from rotational invariance: on the largest lattice $32^4$ the propagator
values for momenta with components $k=(0,0,0,2)$ and $k=(1,1,1,1)$ differ 
by less than 3 standard deviations. For the lattices $16^4$ and $24^4$ we 
have found a good agreement within both gauge-fixing algorithms.

\begin{figure*}
\includegraphics[width=0.55\textwidth]{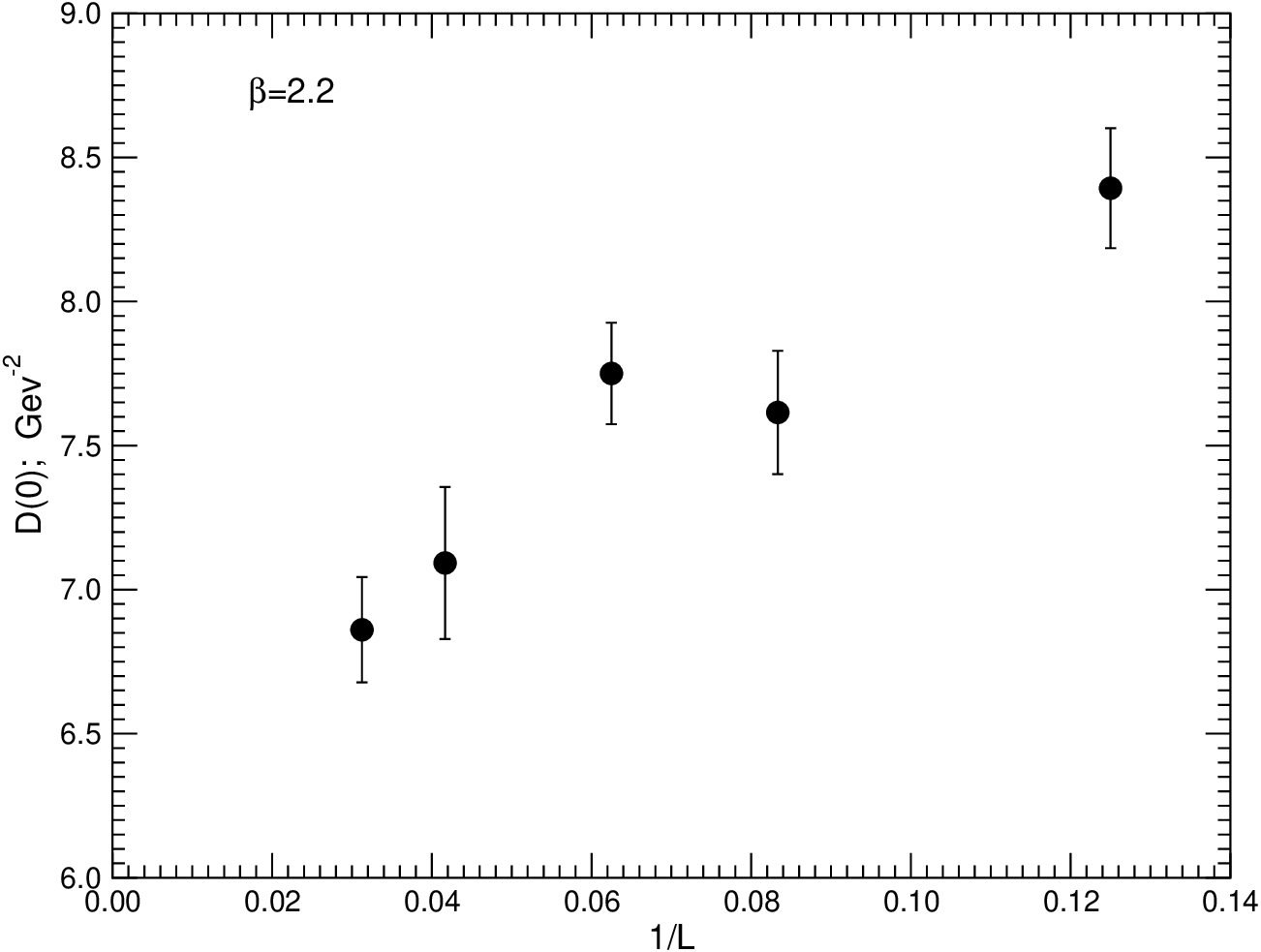}
\caption{Lattice gluon propagator for zero momentum $D(0)$ obtained with the
FSA method as a function of the inverse lattice size.}
\label{fig:Gl_D0}
\end{figure*}

\begin{figure*}
\vspace*{1cm}
\includegraphics[width=0.7\textwidth]{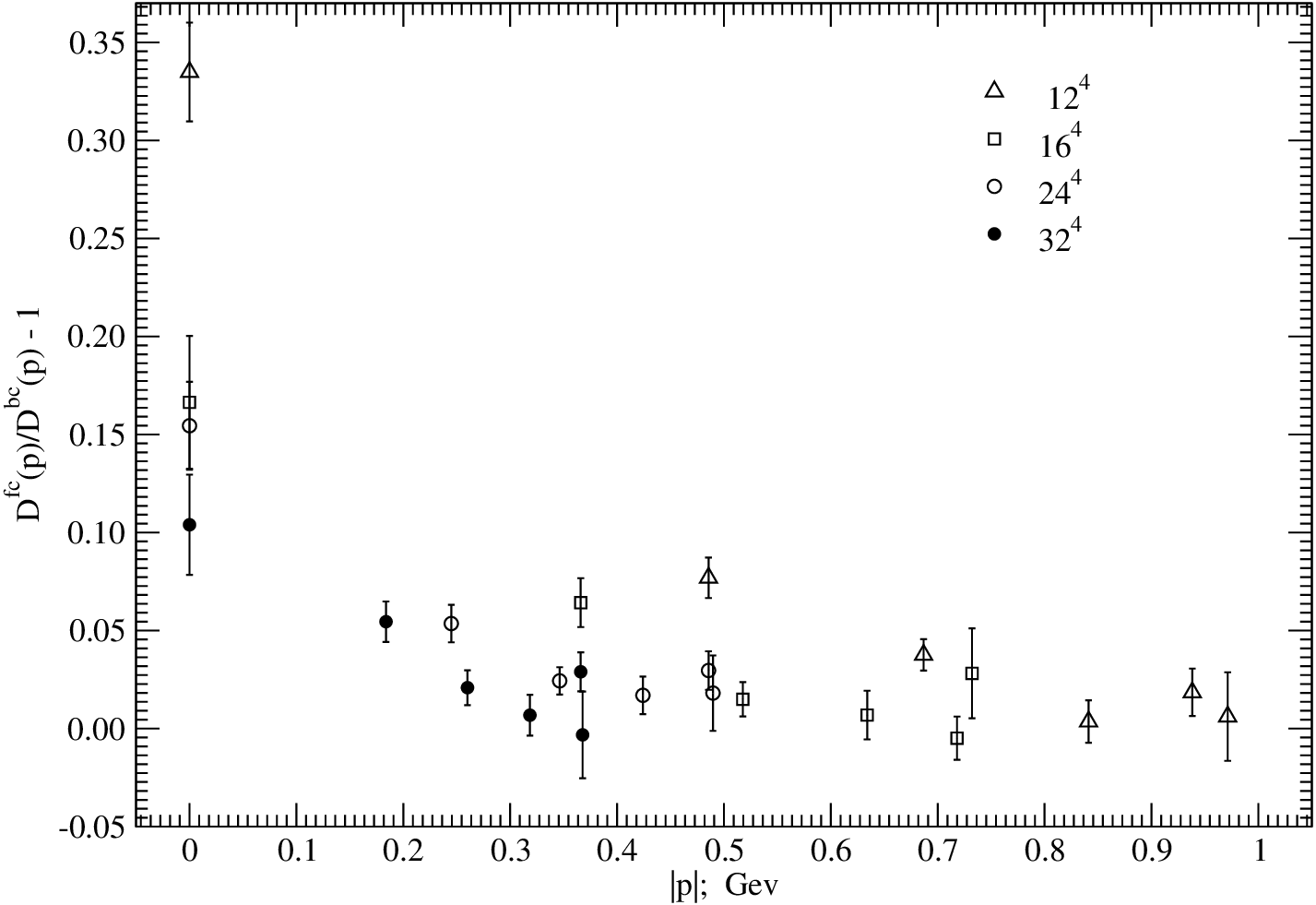}
\caption{The relative difference between the propagator values obtained from
\fc--SA (with one copy in one random flip sector) and obtained from 
\bc--FSA as a function of the momentum $p$ for various lattice sizes.}
\label{fig:change_vs_V_p}
\end{figure*}

In Figs. \ref{fig:SA_vs_OR} and \ref{fig:Gl_main} the data obtained with FSA on the
$32^4$ lattice show a tendency to decrease toward smaller values at the
smallest nonzero momentum. This is the first lattice result in favor of a
decreasing gluon propagator towards the infrared in four dimensions.
In both figures we have also shown the values of the gluon propagator
at zero momentum, $D(0)$, which has a monotonous downward volume dependence
(compare also Fig.~\ref{fig:Gl_D0})~\cite{Boucaud:2006pc}.
However, the value of $D(p \equiv 0)$ is expected to be affected by
stronger finite-volume (and Gribov ambiguity) effects than $D(p_{\rm min} 
\to 0)$~\cite{Fischer:2007pf}.
We have also checked, whether our result can be seen in agreement with
the expectation  $~D(p~\to~0)=0$. Indeed, a fit restricted to the interval 
$ 0 < p < 500$ MeV with the function
\beq
D(p)=p^{\,2 \alpha} \cdot (g_0 + g_1 \cdot p^2 ) \; , 
\eeq
worked perfect ($\chi^2/{\rm d.o.f.}=0.06$) with an exponent $\alpha = 0.09(1)$, 
which is in qualitative agreement with the DSE result \cite{Lerche:2002ep}
$\kappa_D \equiv 1 + \alpha = 1.19$.
Although this cannot be taken too seriously, 
our result gives some credit to the assumption that we are beginning to see 
the gluon propagator to decrease toward zero momentum.
The replacement of the \fc--SA algorithm (i.e. with one copy $nc=1$ in one 
random flip sector $ns=1$) by the \bc--FSA algorithm (with $16$ sectors under 
control and the preferential strategy according to Table~\ref{tab:statistics}) 
leads to a systematic change of the resulting propagator which is
presented in Fig.~\ref{fig:change_vs_V_p}.
The Figure shows for all lattice volumes that for fixed lattice size the
relative deviation of the FSA results for the gluon propagator from the
simple SA results decreases with increasing momentum going rather quickly to
zero within error bars. Furthermore, for fixed physical momentum the
relative deviation goes to zero with increasing volume, indicating that
the two Landau gauge-fixing prescriptions (without and with flips) become
equivalent in the large volume limit. On the other hand, if we compare data
for the minimal momenta for every lattice we find that the respective relative
deviation decreases with increasing lattice volume rather slowly indicating that
the effect of flip sectors for the minimal momentum will be important for all
accessible lattices. This is also a valid conclusion, although to a smaller
extent, for the next-to-minimal momentum.

\section{Conclusions}
In this paper we have reinvestigated the Landau gauge gluon propagator
on the lattice within $SU(2)$ pure Yang-Mills theory. Our main achievement
is the use of an improved gauge-fixing prescription which takes into account
$\mathbb{Z}(2)$ flip transformations equivalent to nonperiodic gauge
transformations as well as the use of the simulated annealing method in combination
with subsequent overrelaxation steps. Comparing with the exclusive use of standard
overrelaxation without applying flips we confirm clear Gribov copy effects for the
gluon propagator. But more important, we observe that finite-size effects seem to 
become suppressed for a gauge-fixing prescription providing copies closer to the 
fundamental modular region. For the first time in the 4d $SU(2)$ case on symmetric
lattices we see a flattening or a signal for a turnover giving access to a limit
$D(q \to 0)=0$ in agreement with DSE predictions and confinement scenarios by
Zwanziger~\cite{Zwanziger:1993dh}  or Kugo and Ojima~\cite{Kugo:1979gm}

\section*{ACKNOWLEDGEMENTS}
V.~G.~B., E.-M.~I, and M.~M-P. wish to thank Boris Martemyanov for useful 
remarks and discussions. G.~B. is grateful to Jan M. Pawlowski and Holger 
Gies for their interest and discussions.

This investigation has been partly supported by the Heisenberg-Landau
program of collaboration between the Bogoliubov Laboratory of Theoretical 
Physics of the Joint Institute for Nuclear Research Dubna (Russia) and 
German institutes and partly by the joint DFG-RFBR grant 436 RUS 113/866/0-1
and the RFBR-DFG grant 06-02-04014.
V.~G.~B. and V.~K.~M. acknowledge support by the RFBR grant 05-02-16306,
and V.~G.~B. is presently supported by the RFBR grant 07-02-00237.
G.~B. acknowledges support from DFG grants Re856/6-1 and Re856/6-2.
M.~M.-P. and E.-M.~I. appreciate the support from DFG under the grant 
FOR 465 / Mu932/2-2.


\end{document}